\documentclass[10pt,prl,twocolumn,floatfix]{revtex4}
\usepackage{amsfonts}
\usepackage{amssymb}
\usepackage{amsmath}
\usepackage[dvips]{graphicx}

\usepackage{amsmath}
\usepackage{graphicx}
\usepackage{amsfonts}
\usepackage{amssymb}
\usepackage{hyperref}
\usepackage{color}
\usepackage{mathrsfs}
\usepackage{isomath}
\usepackage{amsmath}
\usepackage{amsthm}
\usepackage{epstopdf}
\usepackage{txfonts}

\begin{document}

\title{Experimental demonstration of robustness of Gaussian quantum coherence}
\author{Haijun Kang$^{1,2}$}
\author{Dongmei Han$^{1,2}$}
\author{Na Wang$^{1,2}$}
\author{Yang Liu$^{1,2}$}
\author{Shuhong Hao$^{3}$}
\email{haoshuhong@qq.com}
\author{Xiaolong Su$^{1,2}$}
\email{suxl@sxu.edu.cn}
\affiliation{$^{1}$State Key Laboratory of Quantum Optics and Quantum Optics Devices, \\
Institute of Opto-Electronics, Shanxi University, Taiyuan, 030006, People's
Republic of China\\
$^{2}$Collaborative Innovation Center of Extreme Optics, Shanxi University,\\
Taiyuan,Shanxi 030006, People's Republic of China\\
$^{3}$School of Mathematics and Physics, Anhui University of Technology, Maanshan, 243000, People's Republic of China}

\begin{abstract}
	Besides quantum entanglement and steering, quantum coherence has also been identified as a useful quantum resource in quantum information. It is important to investigate the evolution of quantum coherence in practical quantum channels. In this paper, we experimentally quantify the quantum coherence of a squeezed state and a Gaussian Einstein-Podolsky-Rosen (EPR) entangled state transmitted in Gaussian thermal noise channel, respectively. By reconstructing the covariance matrix of the transmitted states, quantum coherence of these Gaussian states is quantified by calculating the relative entropy. We show that quantum coherence of the squeezed state and the Gaussian EPR entangled state is robust against loss and noise in a quantum channel, which is different from the properties of squeezing and Gaussian entanglement. Our experimental results pave the way for application of Gaussian quantum coherence in lossy and noisy environments.
\end{abstract}
	
	\maketitle
	
	\section{I. Introduction}
	
	The principle of coherent superposition of waves plays important roles in many well-known phenomena, such as interference and diffraction.
	In quantum mechanics, the superposition principle which is one of the fundamental nonclassical characteristics of quantum states, underlies many nonclassical properties of quantum mechanics including entanglement or coherence \cite{Baumgratz2014}. Recently, resource theories of coherence have attracted a lot of attention \cite{Winter2016,Streltsov2017,Chitambar2019}.
	Quantum coherence, which characterizes the quantumness and underpins quantum correlations in quantum systems, plays a key role in many novel quantum phenomena and has been identified as a quantum resource for quantum information processing \cite{Aberg2014,Mark2016,Li2016,Shi2017,Shi2020}. Quantum coherence also plays a strong role in biology systems \cite{Plenio2013}, such as photosynthetic energy transport, the avian compass and sense of smell.
	
	To quantify coherence, Baumgratz et al. established a framework by referring to the method of quantifying entanglement \cite{Baumgratz2014}.
	The quantum coherence of a quantum state is defined as the minimum distance between the quantum state and an incoherent state in the Hilbert space \cite{Baumgratz2014}. In addition to relative entropy and $l_{1}$-norm \cite{Baumgratz2014}, it has been shown that quantum coherence can also be quantified by Fisher information \cite{Feng2017}, skew information entropy \cite{Yu2017}, Tsallis relative $\alpha$ entropy \cite{Rastegin2016}, robustness \cite{Napoli2016}, and so on.
	The frozen \cite{Bromley2015}, distillation \cite{Chitambar2016116}, catalytic \cite{Bu2016}, and erasure \cite{Singh2015} of quantum coherence and the relationships between quantum coherence and complementarity relation\cite{Cheng2015}, uncertainty relation\cite{Yuan2017U}, quantum entanglement or other types of quantum correlation \cite{Xi2015,Chitambar2016117} have also been investigated.
	With the rapid development in quantum coherence theory, the experimental demonstration related to quantum coherence is in progress \cite{Yuan2017D,Wu2017,Gao2018,Wu2018,Zhang2019,Xu2020}. 
	Gaussian states, such as the squeezed state and the Einstein-Podolsky-Rosen (EPR) entangled state, play essential roles in continuous variable (CV) quantum information \cite{Braunstein2005,Wang2007,Weedbrook2012}, where Gaussian states are generated deterministically and information is encoded in the position or momentum quadrature of photonic harmonic oscillators. For example, Gaussian states has been applied in quantum computation \cite{Ukai2011,Su2013}, quantum key distribution \cite{Su2009,Gehring2015,Diamanti2016}, quantum teleportation \cite{Furusawa1998,Huo2018}, quantum entanglement swapping \cite{Takei2005,Jia2004,Su2016}, quantum dense coding \cite{Li2002,Jun2005}, and verification of the error-disturbance uncertainty relation \cite{Liu2019npj,Liu2019pr}. Recently, it has been shown that quantum coherence with infinite-dimensional systems can be quantified by relative entropy \cite{Fan2016}. Then, the investigations of Gaussian quantum coherence attracted lots of attention \cite{Xu2016, Buono2016, Albarelli2017}. 
	
	In practical quantum computation and quantum information, decoherence coming from the inevitable interaction between a quantum resource and the environment is a main obstacle \cite{Siena2005}. Up to now, the decoherence of squeezing, entanglement and quantum steering in the thermal noise channel have been experimentally demonstrated \cite{Barbosa2010,Deng2016,Deng2017,liu2020,Deng2021}. Toward the application of quantum coherence, it is necessary to investigate the evolution of quantum coherence in lossy and noisy environment \cite{Suciu2016,Croitoru2020}. Very recently, it has been shown that quantum coherence can be robust against noise theoretically \cite{Buruaga2017} and quantum coherence of optical cat states can be robust against loss  \cite{Zhang2021}.
	
	In this paper, we experimentally quantify quantum coherence of Gaussian states, for example, a squeezed state and a Gaussian EPR entangled state, by measuring their covariance matrices. Then we investigate quantum coherence of these Gaussian states through lossy and noisy channels. We show that quantum coherence of these Gaussian states is robust against loss in a lossy channel, which is similar to the case of squeezing and entanglement of Gaussian state. The most interesting thing is that the quantum coherence of these Gaussian states is still robust in a noisy channel even if the squeezing and entanglement of Gaussian state disappear. The presented results provide useful reference for applying quantum coherence of a Gaussian state in practical quantum information processing.
	
	\section{II. The principle of experiment}
		
	\begin{figure}[]
		\centering
		\includegraphics[width=0.5\textwidth]{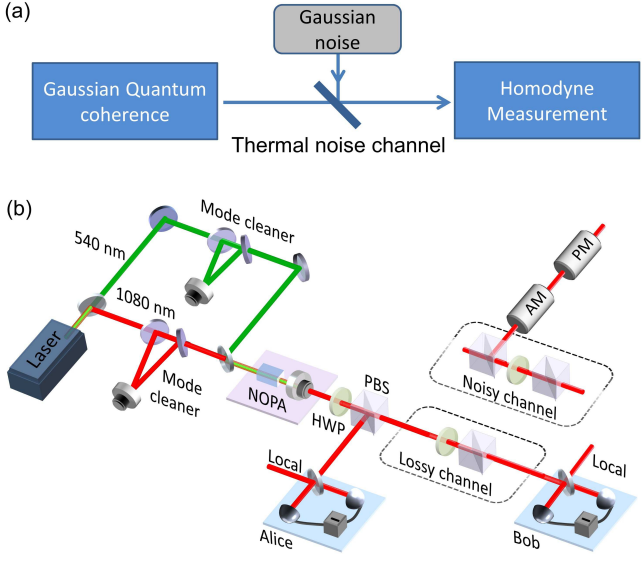}
		\caption{(a) Schematic of transmitting the quantum coherence
			of a Gaussian state in a thermal noise channel. (b) Experimental setup. The 1080~nm and the 540~nm laser output from the Nd:YAP/LBO laser, pass through two mode cleaners and are injected into the nondegenerate-optical-parametric-amplifier (NOPA) as signal light and pump light, respectively. The output modes of the polarization beam splitter (PBS) behind the NOPA are an amplitude squeezed state (transmitted mode) and a phase squeezed state (reflected mode) or EPR entanglement state, when the half wave plate (HWP) behind the NOPA is set to $22.5^\textup{o}$ or $0^\textup{o}$, respectively. We use homodyne detectors to measure the output modes and a digital storage oscilloscope to record the experimental data. The interference efficiencies of homodyne detectors are 99\% and the quantum efficiencies of photodiodes (LASER COMPONENTS, InGaAs-PD-500um) are 99.6\%. AM, amplitude modulator; PM, phase modulator.}
		\label{fig:1}
	\end{figure}

	As shown in Fig.~1(a), the quantum coherence of a Gaussian state is distributed through a Gaussian thermal noise channel, and the quantum coherence of the output state is measured. Here we consider two kinds of Gaussian channel: one is a lossy channel where only vacuum noise is involved, and the other is a noisy channel where the noise higher than vacuum noise exists. The squeezed state and the EPR entangled state, which are generated from a nondegenerate-optical-parametric-amplifier (NOPA) as shown in Fig.~1(b), are used as two examples to investigate the quantum coherence of a Gaussian state in our experiment.
	The NOPA cavity is in a semi-monolithic structure, which is composed by an $\alpha$-cut type-II potassium titanyl phosphate (KTP) crystal ($3\times 3\times 10$ mm),  whose front surface is used as input mirror, and a concave mirror with curvature radius of 50~mm. The NOPA is operating in the case of deamplification, i.e., the relative phase between the signal and the pump light is locked to $(2n+1)\pi$. The lossy channel is simulated by combination of a half-wave plate (HWP) and a polarization beam splitter (PBS) and the noisy channel is simulated by combination of a HWP, two PBSs, and an ancillary coherent beam carrying Gaussian noise.
	
	At first, we distribute the amplitude squeezed state through a lossy channel, where the output state is measured by the homemade homodyne detector at Bob's side in time domain. Then we investigate quantum coherence of an EPR entangled state in lossy channel, where output states are measured by Alice's and Bob's homodyne detectors simultaneously in the time domain. Finally, we quantify the quantum coherence of the amplitude squeezed state and EPR entangled state in a noisy channel, where the Gaussian noise is added through amplitude and phase modulators (LINOS, LM0202 P and LM0202 PHAS) on an ancillary coherent beam and coupled into the lossy channel by the PBS. In the measurement of the covariance matrix of output states in the time domain, the electrical signal of each homodyne detector is mixed with a 3~MHz reference signal (SRS, DS345), then pass through a low-pass filter and a low-noise preamplifier (SRS, SR560), and finally recorded in a digital storage oscilloscope (TELEDYNE LECROY, WaveRunner 640Zi). The bandwidth of the homodyne detectors we used is 8~MHz, and common-mode rejection ratio is 30~dB (at 3~MHz). The sampling rate of the digital storage oscilloscope is 500~KS/s, and there are $5\times 10^5$ data for each sampling space.
	
	\section{III. Quantum coherence}
	
	Quantum coherence of a quantum state $\hat{\rho}$ in Fock space can be calculated by \cite{Baumgratz2014} 
	\begin{equation}
	\mathcal{C}_{rel.~ent.}(\hat{\rho} ) =S\left(\hat{\rho}_{diag}\right)-S\left( \hat{\rho} \right),
	\end{equation}
	where $S$ is van Neumann entropy and $\hat{\rho}_{diag}$ is a diagonal matrix which removing all off-diagonal elements of the density matrix $\hat{\rho}$.
	In the case of Gaussian quantum information, a Gaussian state  $\hat{\rho}(\bar{\mathbf{x}},\mathbf{V})$ can be completely represented by the displacement $\bar{\mathbf{x}}$ and the covariance matrix $\mathbf{V}$ in phase space, which correspond to the first and second statistical moments of the quadrature operators respectively \cite{Wang2007,Weedbrook2012}. The displacement $\mathbf{\bar{x}}=\langle 
	\mathbf{\hat{x}}\rangle $, where	
	$\mathbf{\hat{x}=(}\mathit{\hat{X}}_{\mathit{1}}\mathit{,\mathit{\hat{Y}}_{%
			\mathit{1}},...,\mathit{\hat{X}}}_{\mathit{\mathit{N}}}\mathit{,\hat{Y}}_{%
		\mathit{N}}\mathbf{)}^{\mathit{t}},\mathit{\hat{X}}_{\mathit{k}}=(\hat{a}_{_{k}}+\hat{a}%
	_{_{k}}^{\dagger })$ and $\mathit{\hat{Y}}_{\mathit{k}}=i(\hat{a}%
	_{_{k}}^{\dagger }-\hat{a}_{_{k}})$ are the amplitude and phase quadratures of an optical mode, respectively.
	The element of covariance matrix $\mathbf{V}$ is defined as
	$\mathbf{V}_{ij}=\frac{1}{2}\langle \hat{x}_{i}\hat{x}_{j}+\hat{x}_{j}\hat{x}_{i}\rangle -\langle \hat{x}_{i}\rangle \langle \hat{x}_{j}\rangle$. The diagonal Gaussian states (incoherent states) are thermal states \cite{Xu2016}, so the incoherent state $\hat{\rho}_{diag}$ in Eq. (1) is replaced by an $N$-mode thermal state $\hat{\rho}(\bar{\mathbf{x}}_{th}, \mathbf{V}_{th})$ whose mean number of particle is the same to $\hat{\rho}(\bar{\mathbf{x}},\mathbf{V})$ for each mode. Thus, the Gaussian quantum coherence of an $N$-mode Gaussian state can be represented as \cite{Xu2016}
	\begin{equation}
	\mathcal{C}_{rel.~ent.}\left[\hat{\rho}(\bar{\mathbf{x}},\mathbf{V})\right] =S\left[\hat{\rho}(\bar{\mathbf{x}}_{th},\mathbf{V}
	_{th})\right] -S\left[\hat{\rho}(\bar{\mathbf{x}},\mathbf{V}) \right],
	\end{equation}
	where $S\left[\hat{\rho}(\bar{\mathbf{x}},\mathbf{V})\right]\!=-\!\underset{i=1}{\overset{N}{\sum }}\!\left[\!\left(
	\frac{\nu_{i}-1}{2}\right)\!\log\!_{2}\!\left( \frac{\nu_{i}-1}{2}
	\right)\!-\!\left( \frac{\nu_{i}+1}{2}\right)\!\log\!_{2}\!\left( \frac{\nu
		_{i}+1}{2}\right)\!\right]$ and
	$S\left[ \hat{\rho}(\bar{\mathbf{x}}_{th},\mathbf{V} _{th})\right]\!\!=\!-\!\underset{i=1}{\overset{N}{\sum }}\!\left[\!\left(
	\frac{\mu_{i}-1}{2}\right)\!\log\!_{2}\!\left( \frac{\mu_{i}-1}{2}
	\right)\!-\!\left( \frac{\mu_{i}+1}{2}\right)\!\log\!_{2}\!\left( \frac{\mu_{i}+1}{2}\right)\!\right]$ are the von-Neumann entropy of $\hat{\rho}(\bar{\mathbf{x}},\mathbf{V})$ and $\hat{\rho}(\bar{\mathbf{x}}_{th},\mathbf{V}
	_{th})$, respectively.  
	$\nu_{i}$ and $\mu_{i}$ are the symplectic eigenvalues of $\mathbf{V}$
	and $\mathbf{V}_{th}$, respectively. Here the displacements $\mathbf{\bar{x}}_{th}=0$ and the elements of the diagonal covariance matrix $\mathbf{V}_{th}$ are given by $\mathbf{V}$ with $V_{th}\ _{2i-1,2i-1}=V_{th}\ _{2i,2i}=\frac{1}{2}\left(V_{2i-1,2i-1}+V_{2i,2i}+\left[ \mathbf{\bar{x}}_{2i-1}\right] ^{2}+\left[\mathbf{\bar{x}}_{2i}\right] ^{2}\right)$.
	
	Since the displacements $\bar{\mathbf{x}}$ of the Gaussian states we used in our experiment are zero, the Gaussian state can be completely represented by its covariance matrix $\mathbf{V}$. The covariance matrix of the amplitude squeezed state is given by
	\begin{align}
	\mathbf{V}_{squ}=\left(
	\begin{array}{cc}
	V_s & 0 \\
	0 & V_{as}%
	\end{array}%
	\right),
	\end{align}
	where $V_s$ and $V_{as}$ are the variances of squeezed and antisqueezed noise of the squeezed state, respectively. The squeezed and antisqueezed noise levels of the squeezed state are quantified by $10\log _{10}V_s$ dB and $10\log _{10}V_{as}$ dB, respectively. The symplectic eigenvalue of the squeezed state can be determined by  $\sqrt{\textup{Det}\mathbf{V}_{squ}}$.
	
	The covariance matrix of the EPR entangled state is given by
	\begin{align}
	\mathbf{V}_{ent}&=\left(
	\begin{array}{cccc}
	\mathbf{A} & \mathbf{C} \\
	\mathbf{C}^{t} & \mathbf{B}
	\end{array}%
	\right),
	\end{align}
	where $\mathbf{A}=\mathbf{B}=\frac{1}{2}(V_s+V_{as})~\mathbf{I}$, $\mathbf{C}=\frac{1}{2}(V_s-V_{as})~\mathbf{Z}$, $t$ denotes transpose, 
	$\mathbf{I}=\begin{pmatrix}	
	\begin{smallmatrix}	
	1 & 0 \\
	0 & 1	
	\end{smallmatrix}	
	\end{pmatrix}$ and
	$\mathbf{Z}=\begin{pmatrix}	
	\begin{smallmatrix}	
	1 & 0 \\
	0 & -1	
	\end{smallmatrix}	
	\end{pmatrix}$. The symplectic eigenvalues can be determined by  $\sqrt{\frac{\Delta \pm \sqrt{\Delta^2-4\textup{Det}\mathbf{V}_{ent}}}{2}}$, where $\Delta=\textup{Det}\mathbf{A}+\textup{Det}\mathbf{B}+2\textup{Det}\mathbf{C}$. The positive partial transposition (PPT) criterion \cite{Simon2000} is applied to describe the entanglement of the EPR entangled state, which is a sufficient and necessary condition for a two-mode entanglement state with continuous variables. The PPT value can be determined by  $\sqrt{\frac{\Gamma-\sqrt{\Gamma^2-4\textup{Det}\mathbf{V}_{ent}}}{2}}$, where $\Gamma=\textup{Det}\mathbf{A}+\textup{Det}\mathbf{B}-2\textup{Det}\mathbf{C}$. When the PPT value is less than 1, the two mode quantum states are entangled \cite{Simon2000}. 
	
	A one-mode Gaussian state $\rho (\mathbf{\bar{x},V})$ transmitting in a Gaussian channel can be represented by \cite{Holevo1999,Eisert2005}
	\begin{align}
	\begin{aligned}
	\mathbf{\bar{x}}\rightarrow &T\mathbf{\bar{x}}+\bar{d},\\
	\mathbf{V}\rightarrow &T\mathbf{V}T^{t}+\varLambda,
	\end{aligned}
	\end{align}
	where $\bar{d}$ is displacement operator in phase space, $T$ is amplification or attenuation and rotation operator in phase space, and $\varLambda$ is a noise term that may consist of quantum as well as classical noise. 
	The thermal noise channel, which belongs to incoherent channel \cite{Xu2016}, of CV Gaussian quantum system can be written as \cite{Eisert2005}
	\begin{align}
	\begin{aligned}
	T&=\sqrt{\eta}~\mathbf{I},\\
	\varLambda&=\left(1-\eta\right)\left( \delta+\upsilon\right)~\mathbf{I},\\
	\bar{d}&=0,
	\end{aligned}
	\end{align}
	where $\eta$ and $\delta$ are the transmission efficiency and the excess noise of the Gaussian channel, respectively, and $\upsilon=1$ represents the vacuum noise. When $\delta=0$, the Gaussian channel is a lossy channel. The loss of the channel is given by $L=1-\eta$.
	
	\section{IV. Results}
	
	First, we quantify the quantum coherence of the amplitude squeezed state and the EPR entangled state by the covariance matrices we reconstructed (see Appendix A) in a lossy environment. The decoherence of the squeezing of the squeezed state and the entanglement of the EPR entangled state in the lossy channel is shown in Fig.~2(a) and 2(b), respectively. It is obvious that the squeezing and entanglement are robust against loss in lossy channel.
	Quantum coherence of the squeezed state and the EPR entangled state in lossy channel is shown in Fig.~2(c) and 2(d), respectively. We can find that the quantum coherence of the squeezed state and the EPR entangled state are both decreased with the increase of loss, which reaches zero only when the maximal loss is reached. When the loss equals 1, the squeezed state turns into a vacuum state and the EPR state turns into a separable state. We can see that quantum coherence of these Gaussian states is also robust against loss in a lossy channel.
	
	\begin{figure}[]
		\centering
		\includegraphics[width=0.48\textwidth]{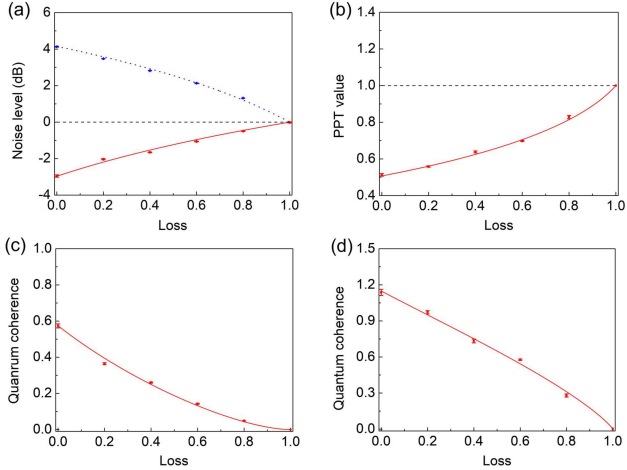}
		\caption{Experimental results in a lossy channel. (a) Dependence of squeezing (red solid line) and antisqueezing (blue dotted line) of the squeezed state on the loss. The dashed line is the shot noise limit (SNL). (b) Dependence of PPT value of the EPR entangled state on the loss. The dashed line is the boundary of entangled and separable state. (c) and (d), Dependence of quantum coherence of the squeezed state and the EPR entangled state on the loss, respectively. The initial squeezed and antisqueezed noise levels are -2.95 dB and 4.15 dB, respectively. The error bars represent one standard deviation and are obtained based on the statistics of the data.}
		\label{fig3}
	\end{figure}
	
	Then we quantify the quantum coherence of a squeezed state and an EPR entangled state in a noisy environment. 
	In the case of noisy channel, we fix the channel losses to $L=0.4$, and change the excess noise $\delta$ added on the amplitude squeezed state and one mode of the EPR entangled state.
	The decoherence of the squeezing of the squeezed state and entanglement of the the EPR entangled state in the noisy channel is shown in Fig.~3(a) and 3(b), respectively. Different from the results in the lossy channel, the noise level of squeezing is beyond the shot noise limit (SNL) when the excess noise overs $\delta=0.74$ and the PPT value is greater than 1 when the excess noise is greater than $\delta=2.14$, which means the squeezing of the squeezed state and the entanglement of the EPR entanglement state are destroyed.
	
	\begin{figure}[]
		\centering
		\includegraphics[width=0.48\textwidth]{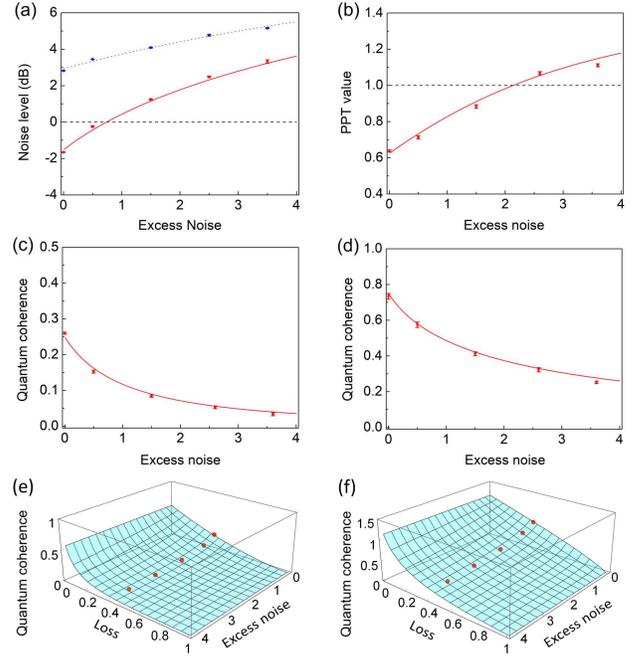}
		\caption{Experimental results in a noisy channel. (a) Dependence of squeezing (red solid line) and antisqueezing (blue dotted line) of the squeezed state on the excess noise. The dashed line is SNL. (b) Dependence of PPT value of the EPR entangled state on the excess noise. The dashed line is the boundary of entangled and separable state. (c) and (d), Dependence of quantum coherence of the squeezed state and the EPR entangled state on the excess noise, respectively. (e) and (f), Quantum coherence of the squeezed state and the EPR entangled state parameterized by loss and excess noise, respectively. The red dots represent the experimental results in (c) and (d).}
		\label{fig4}
	\end{figure}
	
	The quantum coherence of the squeezed state and the EPR entangled state in a noisy channel is shown in Fig.~3(c) and 3(d), respectively. It is obvious that the quantum coherence of the squeezed state and the EPR entangled state is both decreased with the increase of the excess noises. However, the quantum coherence of these Gaussian states still exists even if the squeezing and entanglement disappear. It is interesting that quantum coherence of these two Gaussian states will vanish only when infinite excess noises are involved in the case of fixed channel loss. The dependence of quantum coherence of the squeezed state and the EPR entangled state on loss and excess noise is shown in Fig. 3(e) and 3(f), respectively. It is obvious that the quantum coherence of these Gaussian states is robust against noise in a noisy channel.
	
	We note that the squeezing and entanglement are destroyed at different excess noise levels as shown in Fig.~3(a) and 3(b). The reason for this result is that the excess noise is only added on one mode of the EPR entangled state. It shows that the EPR entanglement can tolerate more noise than squeezed state in one-mode noise channel. The squeezing and entanglement will be destroyed at same excess noise level if we add the excess noise on both modes of the EPR entangled state simultaneously (see Appendix B).
	
	We also demonstrate the monotonicity of quantum coherences of the squeezed state and the entanglement state in lossy and noisy channels as shown in Fig.~2 and 3, respectively. The quantum coherences of these two Gaussian states are decreasing with the increase of loss and noise in quantum channels, which is because the lossy and noisy channels are all incoherent operations and quantum coherence will decrease under incoherent operations \cite{Baumgratz2014,Xu2016}. The physical reason for the robustness of quantum coherences of these Gaussian states in a noisy channel is that the proportion of quantum coherence is decreased when it is mixed with thermal noise, but the quantum coherence disappears completely only when infinite thermal noise is involved.
	
	\section{V. Conclusion}
	
	In summary, we experimentally demonstrate the quantum coherence of Gaussian states in lossy and noisy channel. 
	The results confirm that quantum coherences of the squeezed state and the EPR entangled state are robust against loss and noise in a Gaussian thermal noise channel, although the squeezing and entanglement of a Gaussian states disappear at a certain noise level in a noisy channel. Thus, the quantum coherence of Gaussian state can resist the decoherence when it is used as quantum resource. Our investigation makes a step toward the application of quantum coherence as a quantum resource in quantum communication.
	
	It is interesting to accomplish quantum information tasks that only require quantum coherence of a Gaussian state due to its unique property of it in the presence of loss and noise. However, a suitable application for only applying the quantum coherence of a Gaussian state remains an open question. Recently, it has been shown that the Gaussian entanglement can be transferred in a single-mode cavity \cite{Bougouffa2020}, which is an application of robustness of quantum coherence of Gaussian states. Based on the presented results of robustness for quantum coherences of Gaussian states, the potential application of quantum coherence is worthy of further investigation.

	\appendix
	\setcounter{figure}{0}
	\renewcommand{\thefigure}{B\arabic{figure}}
	\medskip
	\section*{APPENDIX A: Reconstruction of the covariance matrix}
	To reconstruct the covariance matrix of an EPR entangled state, the variances and the cross correlations of the amplitude or phase quadratures are obtained by simultaneously measuring the amplitude or phase quadratures of two modes of the EPR entangled state in the time domain. The diagonal elements of the covariance matrix are the variances of the amplitude and phase quadratures $\langle \Delta^2 (\hat{x}_i) \rangle$, and the nondiagonal elements are the covariances of the amplitude or phase quadratures, which are calculated via the measured variances \cite{Steinlechner2013}
	\begin{equation}
	V_{ij}=[\langle \Delta^2 (\hat{x}_i+\hat{x}_j) \rangle-\langle \Delta^2 (\hat{x}_i) \rangle-\langle \Delta^2 (\hat{x}_j) \rangle]/2,
	\end{equation}
	\begin{equation}
	V_{ij}=-[\langle \Delta^2 (\hat{x}_i-\hat{x}_j) \rangle-\langle \Delta^2 (\hat{x}_i) \rangle-\langle \Delta^2 (\hat{x}_j) \rangle]/2,
	\end{equation}
	where $\langle \Delta^2 (\hat{x}_i-\hat{x}_j) \rangle$ and $\langle \Delta^2 (\hat{x}_i+\hat{x}_j) \rangle$ are the correlation variances of amplitude and phase quadratures, which can be obtained from the measured variances in the time domain. Based on the reconstructed covariance matrix, the quantum coherence of the EPR entangled state can be quantified according to Eq. (2).
	
	\section*{APPENDIX B: Two Gaussian thermal noise channels}
	Here, we consider the situation in which two modes of the Gaussian EPR entangled state are distributed through two Gaussian thermal noise channels, where the loss and excess noise added on both modes of EPR entangled state, as shown in Fig.~B1. The two output states are measured by Alice's and Bob's homodyne detectors.
	\begin{figure}[h]
		\centering
		\includegraphics[width=0.5\textwidth]{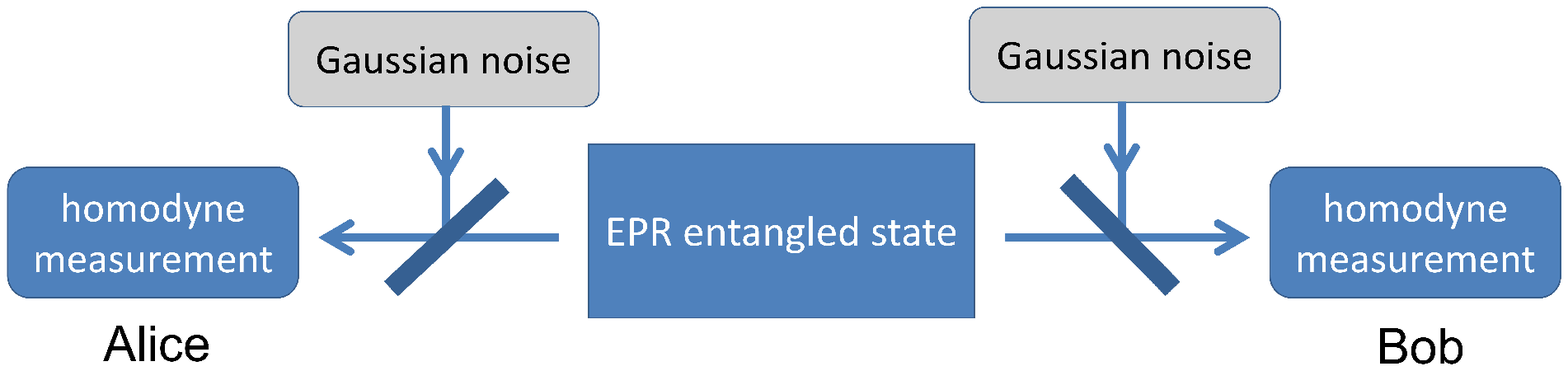}
		\caption{Schematic of transmitting an EPR entangled state in two Gaussian thermal noise channels.}
		\label{fig:A1}
	\end{figure}
	
	The decoherence of the entanglement of the EPR entangled state in the two lossy and noisy channels is shown in Fig.~B2(a) and B2(b), respectively. The entanglement of the EPR entangled state is robust against loss in two lossy channels, where we assume the losses in the two channels are the same. In the case of the two noisy channels, we fix the losses of two Gaussian thermal noise channels to $L=0.4$, and add the same excess noise $\delta$ on the two modes of the EPR entangled state, respectively. 
	The entanglement of the EPR entangled state is destroyed when the excess noise $\delta>0.74$, which is same as the case in which the squeezing of the squeezed state is destroyed as shown in Fig. 3(a) in the main text. Comparing Fig. B2(b) and Fig. 3(b) in the main text, it is obvious that critical point where the disappearance of entanglement happens is different when the excess noise is added on one or two modes of the EPR entangled state. 
	
	The quantum coherences of the EPR entangled state in two lossy and noisy channels are shown in Fig.~B2(c) and B2(d), respectively. The dependence of the quantum coherence of the EPR entangled state on loss and excess noise is shown in Fig. B2(e). It is obvious that quantum coherence is robust against loss and noise when two modes of the EPR entangled state are transmitted through two Gaussian thermal channels respectively, which is the same as the case in which one mode of the EPR entangled state is transmitted through a Gaussian thermal channel as shown in Fig. 3 in the main text.
	
	\begin{figure}[h]
		\centering
		\includegraphics[width=0.48\textwidth]{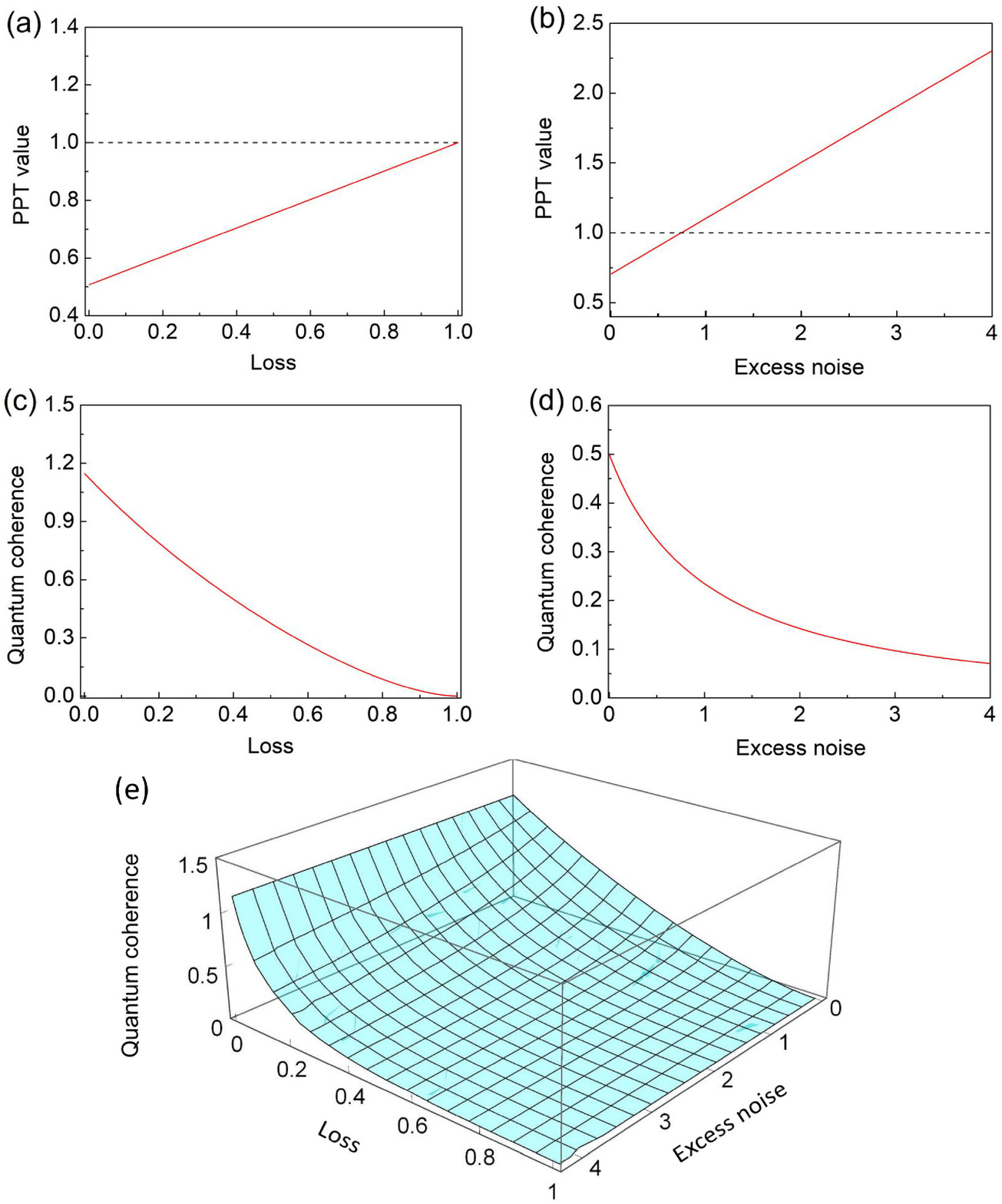}
		\caption{(a) and (b), Dependence of PPT values of the EPR entangled state on the loss and the excess noise, respectively. The dashed line is the boundary of the entangled and separable state. (c) and (d), Dependence of quantum coherence of the EPR entangled state on loss and the excess noise, respectively. (e) Quantum coherence of the EPR entangled state parameterized by loss and excess noise.}
		\label{fig:A2}
	\end{figure}
	
	\medskip
	
	\noindent\textbf{Funding.} National Natural Science Foundation of China (11834010, 11804001); Natural Science Foundation of Anhui Province (Grant Nos 1808085QA11); National Key R\&D Program of China (Grant No. 2016YFA0301402); Fund for Shanxi \textquotedblleft 1331 Project\textquotedblright\ Key Subjects Construction.

\end{document}